\documentclass[12pt]{article} 
\usepackage{graphicx,subfigure,latexsym}

\newcommand{\be}{\begin{equation}}
\newcommand{\ee}{\end{equation}}
\newcommand{\bear}{\begin{eqnarray}}
\newcommand{\eear}{\end{eqnarray}}
\newcommand{\ba}{\begin{array}}
\newcommand{\ea}{\end{array}}

\begin{document}

\begin{titlepage}
\vfill
\begin{flushright}
{\normalsize arXiv:xxxx.xxxx[hep-th]}\\
\end{flushright}

\vfill
\begin{center}
{\Large\bf Chiral helix in AdS/CFT with flavor }

\vskip 0.3in
Dmitri E. Kharzeev$^{1,2}$\footnote{e-mail:
{\tt dmitri.kharzeev@stonybrook.edu}}
 and 
Ho-Ung Yee$^{1}$\footnote{e-mail:
{\tt hyee@tonic.physics.sunysb.edu}}
\vskip 0.15in

 {\it $^{1}$Department of Physics and Astronomy, Stony Brook University,} \\
{\it Stony Brook, New York 11794-3800 }\\[0.3in]
{\it $^{2}$Department of Physics, Brookhaven National Laboratory,} \\
{\it Upton, New York 11973-5000 }\\[0.3in]

{\normalsize  2011}

\end{center}

\vfill

\begin{abstract}
The D3/D7 holographic model aims at a better approximation to QCD by adding to $N=4$ SYM theory $N_f$ of $N=2$ supersymmetric hypermultiplets in the fundamental representation of $SU(N_c)$ -- the ``flavor fields" representing the quarks. 
Motivated by a recent observation of the importance of the Wess-Zumino-like (WZ) term for realizing the chiral magnetic
effect within this model, 
we revisit the phase diagram of the finite temperature, massless D3/D7 model in the presence of external electric/magnetic fields and at finite chemical potential. 
We point out that the A-V-V triangle anomaly represented by the WZ  term in the D7 brane probe action implies the 
existence of new phases that have been overlooked in the previous studies. 
In the case of an external magnetic field and at finite chemical potential, we find a ``chiral helix" phase in which the $U(1)_A$ angle of D7 brane embedding
increases monotonically along the direction of the magnetic field -- this is a geometric realization of the chiral spiral phase in QCD. We also show that in the case of parallel electric and magnetic fields $(E,B)$ there exists a phase in which the 
D7 brane spontaneously begins to rotate, so that the $U(1)_A$ angle changes as a function of time -- this may be called the ``spontaneous rotation" phase; it is a geometrical realization of a phase with non-zero chiral chemical potential.  
Our results call for a more thorough study of the $(T,B,E,\mu)$ phase diagram of the massless D3/D7 model taking a complete account of the WZ term.
 We also speculate about the possible phase diagram in the massive case.

\end{abstract}

\vfill

\end{titlepage}
\setcounter{footnote}{0}

\baselineskip 18pt \pagebreak
\renewcommand{\thepage}{\arabic{page}}
\pagebreak

\section{Introduction}

This work is motivated in part by a recent realization of the chiral magnetic effect (CME) \cite{Kharzeev:2007jp,Fukushima:2008xe} in the D3/D7 model 
via the Wess-Zumino-like (WZ) term in the D7 probe brane action \cite{Hoyos:2011us}\footnote{Closely related phenomena have been discussed in the physics of neutrino emission  \cite{Vilenkin:1979ui}, 
primordial electroweak plasma \cite{Giovannini:1997gp} and quantum
wires \cite{acf}; the separation of electric charge in QCD plasma induced by the chirality imbalance in the presence of magnetic field and/or angular momentum was first discussed in \cite{Kharzeev:2004ey,Kharzeev:2007tn}.}.
The field theory dual to the D3/D7 holographic model is the N=4 supersymmetric Yang-Mills (SYM) theory with $N_f$ of N=2 hypermultiplets in the fundamental representation of the color group $SU(N_c)$ -- the ``flavor fields" representing the quarks.
In this dual field theory, the axial phase of a Dirac quark
in the hypermultiplet is tied to a $U(1)_R$ subgroup of the $SO(6)_R$ symmetry of the N=4 SYM sector.
The hypermultiplet also has its own flavor $U(1)_V$ symmetry which is vector-like to the hypermultiplet quarks.
These ingredients are sufficient to give rise to the $U(1)_V^2 U(1)_R$ flavor anomaly through the triangle one-loop diagram involving the hypermultiplet quarks. In the holographic picture of D7 probe brane in the large $N_c$ D3 brane background,
this triangle anomaly is captured by a WZ--like coupling in the D7 probe brane action.  
A peculiar feature in this set-up is that $U(1)_R$ (we will also call it $U(1)_A$ interchangeably)
is geometrically realized as a $U(1)$ angle in the 5-sphere of $AdS_5\times S^5$ geometry, whereas the 
dynamics of $U(1)_V$ is as usual described by the gauge field on the D7 brane -- therefore the triangle anomaly
is no longer represented by a 5D Chern-Simons term. 

Because $U(1)_R$ is shared by other adjoint matter fields in N=4 SYM, the $U(1)_R$ charges in the hypermultiplet
can be lost to the adjoint sector, and the axial chemical potential can be meaningful only in a quasi-equilibrium sense. 
A novel idea in Ref.\cite{Hoyos:2011us} is to introduce an external time-dependent axial phase into the system that
simulates an axial chemical potential. As $U(1)_A$ is geometrically realized as an angle in the 5-sphere, this
can be achieved by externally rotating the D7 brane along the $U(1)_A$ angle. Considering that the axial charge can be lost
to the adjoint sector, it is a nice way of maintaining a chiral imbalance in an otherwise equilibrated system \cite{Hoyos:2011us}.
This is in fact similar to the true axial $U(1)_A$ symmetry of real QCD where at 
finite temperature the sphalerons can create or annihilate the axial charge. Indeed, a time-dependent QCD $\theta$-angle
which is equivalent to a time-dependent axial phase by anomaly was one way of effectively describing the axial chemical
potential, with $\mu_A = \dot{\theta}$ \cite{Kharzeev:2009fn}. Since in the $\theta$-vacuum picture $\theta$ can be interpreted as a quasi-momentum (in analogy to the Bloch crystal), $\dot{\theta}$ may be interpreted as an external ``force" that maintains an imbalance between the left and right fermions.  This can also be realized holographically, for example
in the Sakai-Sugimoto model, through a time-dependent $C^1_{RR}$ holographic dual to $\theta_{QCD}$ which could clarify the treatment of the holographic chiral magnetic effect \cite{Yee:2009vw,Rebhan:2009vc,Gorsky:2010xu,Rubakov:2010qi,Gynther:2010ed,Brits:2010pw,Kalaydzhyan:2011vx}.

Indeed, one issue with the holographic description of the CME is that in Minkowski signature black hole space-time,
the temporal component of the vector potential $A_\mu$ at the UV boundary is not quite equivalent to the chemical potential even in the case of 
non-anomalous global symmetries \cite{Gynther:2010ed,huy}. The thermal Green's functions in Minkowski signature black hole space-time 
have the structure 
\be
{\rm Tr}_{\rm ensemble}\left( {\cal O}(t) {\cal O}(t')\right)_{\rm time\,\,ordering}\quad,
\ee
where ${\cal O}(t)$ is an operator that evolves as
\be
{\cal O}(t)= e^{+i \hat H t} {\cal O}(0) e^{-i \hat H t}\quad,
\ee
while the ensemble trace can be either grand canonical or (micro) canonical,
\be
{\rm Tr}_{\rm ensemble}= {\rm Tr}\left(e^{-\beta(\hat H-\mu \hat N)}\right) \quad{\rm or}\quad 
{\rm Tr}\left(e^{-\beta\hat H}\right)\quad,
\ee
where $\hat N$ is the charge operator.
A crucial point is that the hamiltonian appears here in two places with rather different roles -- 
the one appearing in the ensemble trace, $e^{-\beta(H-\mu N)}$ or $e^{-\beta \hat H}$, is responsible for the Euclidean imaginary time evolution of period $\beta$
while the one appearing in ${\cal O}(t)$ is responsible for the  evolution in Minkowski real time. In Minkowski black hole space-time,
one sees only the Minkowski evolution and the Euclidean one is not manifest geometrically.
Conversely, in the Euclidean black hole geometry, only the Euclidean evolution is geometrically realized. To be more explicit,
by turning on $A_t(\infty)$ at the UV boundary in Minkowski signature black hole, one achieves a famous replacement
according to the AdS/CFT dictionary,
\be
\hat H\to \hat H-A_t(\infty) \hat N\quad;
\ee
the question however is which hamiltonian should we replace on the field theory side: the one in the ensemble average or the one in ${\cal O}(t)$, or both?
Since the boundary field theory is in Minkowski signature space-time, it seems clear that the hamiltonian in ${\cal O}(t)$ has to be
replaced;  
however 
the question whether the hamiltonian in the ensemble average also has to be replaced is more difficult to answer.
To establish the identification $A_t(\infty)=\mu$ in the ensemble average one should go to the Euclidean signature
black hole where the imaginary time evolution becomes geometrically realized. 

It is clear from the above discussion that
one should not naively associate $A_t(\infty)$ in Minkowski black hole space-time with a chemical potential.
This is relevant because the ambiguities of the chiral magnetic current appear precisely due to $A_t(\infty)$ in Minkowski signature space-time, and are likely 
related to its effect on ${\cal O}(t)$ which should not have been the case for the true chemical potential.
Without $A_t(\infty)$ the ambiguities of the holographic CME disappear \cite{Gynther:2010ed,huy}. 
In Minkowski black hole, the value of the chemical potential can be obtained
by the work done to a unit charge in bringing it from the UV boundary to the black hole horizon.
The issue of singularity at the horizon poses no
problem because what matters in Minkowski dynamics is only the future event horizon and the gauge field is regular there
for any choice of $A_t(\infty)$ \cite{huy}. This is in accord with the fact that $A_t(\infty)$ in Minkowski black hole is not quite
the chemical potential.
One may choose to perform a bulk gauge transformation to remove $A_t$ while introducing a time-dependent $A_r$ instead,
\be
A_r= -t \partial_r A_t(r)\quad.
\ee
Physically, this introduces a time-dependent Wilson line stretching from the UV boundary to the horizon,
\be
W=e^{i\int_{r_H}^\infty A_r}=e^{i t\mu}\quad,
\ee
with frequency which is equal to the chemical potential. What is important here is that there is no singularity
issue with this Wilson line. In fact, the norm $A_M A_N g^{MN}$ one uses in the singularity
argument is not a gauge-invariant concept.

Although one needs an axial chemical potential to observe the CME, there are other 
 interesting phenomena stemming from the triangle anomaly.
One example is the chiral separation effect \cite{Son:2004tq,Metlitski:2005pr}: the emergence of an axial current along the magnetic field $\vec B=B \hat x^3$ in the presence of an ``ordinary" vector chemical potential $\mu_V$, 
\be
j_A^3 = {N_c\over 2\pi^2}\mu_V B\quad.\label{cse}
\ee
The close connection between the CME and the chiral separation effect is particularly easy to see in the case of a strong magnetic field when 
dimensional reduction is appropriate \cite{Basar:2010zd}. In the dimensionally reduced $(1+1)$ theory, the axial and vector currents are related by $j_\mu = \epsilon_{\mu\nu}j^\nu_A$, so that 
the axial charge density $j^0_A$ induces a vector current $j^1$ (CME) and the vector charge density $j^0$ induces an axial current $j_A^1$ (chiral separation).

Contrary to the axial chemical potential, the vector chemical potential (as well as magnetic field) 
is much easier to introduce in the D7 brane. Once we accept (\ref{cse}), it is natural to look for a  signal of the axial current $j_A^3$ in the following way: as $U(1)_A$ is geometrically realized the axial current
will take a form of a chiral spiral \cite{Schon:2000he,Son:2007ny,Basar:2009fg,Bringoltz:2009ym,Kojo:2009ha,Basar:2010zd,Kim:2010pu} that is, the $U(1)_A$ angle of the D7 brane embedding shape should have
a constant gradient along the space direction $x^3\equiv z$.
Even though one is working in the massless limit, the D7 brane
should develop a profile of non-zero axial phase gradient along $x^3$ to satisfy the constraint (\ref{cse}) dictated
by the triangle anomaly. We will call this the chiral helix phase emphasizing its geometrical realization in the holographic setup. 

The phase diagram of D7 brane dynamics in the presence of both magnetic field and ordinary chemical potential
at finite temperature has been studied before \cite{Ammon:2009jt,Evans:2010iy}, but the triangle anomaly constraint (\ref{cse}), or equivalently the effect of the WZ 
term in the D7 brane action, seems to have been missed in these analyses. 
In this paper, we find a significant modification of the phase diagram due to the anomaly, and prove the emergence of the chiral helix phase through a dynamical instability.
A more thorough study of the full $(B,\mu)$ phase diagram will be presented elsewhere \cite{col}.

Another interesting and well-known effect stemming from the triangle anomaly is the creation or annihilation of the $U(1)_A$ charge 
in the presence of parallel electric and magnetic fields, 
\be\label{ax_an}
\partial_\mu j_A^\mu \sim \vec E\cdot \vec B\neq 0\quad.
\ee
At finite chiral chemical potential $\mu_A$, (\ref{ax_an}) determines the power of the chiral magnetic current in the case of the parallel 
electric and magnetic fields:
\be
P = \int d^3 x \ {\vec j} \cdot {\vec E} = \mu_A\ \frac{e^2}{2\pi^2}\ \int d^3 x \ {\vec E} \cdot {\vec B}.
\ee
Note that no power is dissipated in the absence of the electric field, and that depending on the relative signs of $ {\vec E} \cdot {\vec B}$ and $\mu_A$, the power 
can be either positive or negative -- in the latter case the current is powered by the energy stored in the system due to the difference of Fermi energies of left and right 
chiral charges \cite{Fukushima:2008xe,Kharzeev:2009fn}. The reversibility of the sign of the power signals the lack of dissipation that is a consequence of the time reversal invariance of chiral magnetic conductivity and other anomalous transport coefficients -- this provides an important constraint on the anomalous hydrodynamics \cite{Kharzeev:2011ds}.

In the holographic setup one can easily introduce the electric/magnetic fields through world-volume gauge field on the D7 brane.
Because the $U(1)_A$ charge is simply the angular momentum of the D7 brane along the axial angle in the 5-sphere,
the above triangle anomaly constraint implies that the D7 brane should start rotating along this $U(1)_A$ angle
with an increasing speed. Due to the loss of axial charge (dissipation) to the adjoint sector, one expects the system to
stabilize with a finite angular momentum afterwards. We will call this the ``spontaneous rotation phase". 
The existence of this phase seems to have been missed in the previous analyses in the literature \cite{Ammon:2009jt,Evans:2011mu}; 
we expect that the spontaneous rotation phase leads to a significant modification of the phase diagram. 
In the present work we prove the existence of the spontaneous rotation phase by observing a dynamical instability
towards it. The study of the full phase diagram of $(B,E,\mu)$ is deferred to future. 

We stress that there are no externally driven time-dependent parameters in our situations, contrary to Ref.\cite{Hoyos:2011us}.
This is because the chemical potential and electric/magnetic fields of the baryonic $U(1)_V$ symmetry are easier and more natural to introduce through the D7 world-volume
$U(1)_V$ gauge field. Yet we observe that the existence of the $U(1)_V^2 U(1)_R$ triangle anomaly represented by the WZ term
still leads to interesting consequences.

Although we focus only on the massless case in this paper, it is interesting to speculate about the massive case.
In  the case of the chiral spiral in the chiral-symmetry broken phase of QCD, the massive case features an 
inhomogeneous chiral spiral : the axial phase jumps only in  narrow periodic ranges of $x^3$ \cite{Son:2007ny}.
This is due to a competition of the free energy associated with the phase gradient and the anomaly constraint imposed by (\ref{cse}). 
One naturally expects that a similar phenomenon would happen in the massive D3/D7 model, and it will be interesting
to pursue this further.

\section{Massless D3/D7 model with $(T,B,\mu)$}

We study the dynamics of the probe D7 brane embedded in a gravity background of large $N_c$ D3 branes
at finite temperature $T$ given by
\bear
ds^2&=&{r^2\over L^2}\left(-V(r)dt^2 + \sum_{i=1}^3 (dx^i)^2 \right)+{L^2\over r^2 V(r)} dr^2 +L^2 d\Omega_5^2\quad,\quad 
V(r)=1-\left(\pi L^2 T\over r\right)^4\quad,\nonumber\\
F_5^{RR}&=& {(2\pi l_s)^4 N_c\over \pi^3} \epsilon_5\quad,\label{metric}
\eear
where $L^4=4\pi g_s N_c l_s^4\equiv \lambda l_s^4$ and $\epsilon_5$ is the volume form of a unit 5-sphere; $g_s$ and $l_s$ are the string coupling and string length. 
In the weak coupling limit, the D3 branes span $(t,x^i),i=1,2,3$ and the D7 brane wraps additional four dimensions we call
$x^{4,5,6,7}$. The remaining two dimensions $x^{8,9}$ are transverse to both D3 and D7 branes.
The six dimensional space $x^{4-9}$ is the total transverse space to the D3 brane, and 
the radial direction $r$ and the 5-sphere $\Omega_5$ in (\ref{metric}) are the radius and angles in this space.
The theory of the D3/D7 branes is N=4 SYM theory of $SU(N_c)$ plus N=2 hypermultiplet of fundamental representation fields.
In the strong coupling regime that we are interested in, the D3 branes are replaced by the above gravity-flux background (\ref{metric})
whereas the D7 brane is treated as a probe brane to this holographic background. On the field theory side, the dynamics of the probe D7 branes should be dual to the 
dynamics of N=2 hypermultiplet at strong coupling in the quenched approximation.

\begin{figure}[t]
	\centering
	\includegraphics[width=10cm]{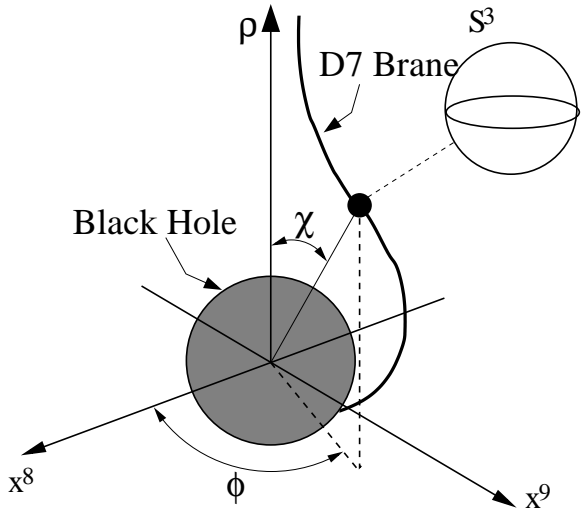}
		\caption{A schematic picture of D7 brane embedding in the D3 background. The $(x^8,x^9)$ plane of our interest is shown explicitly.\label{fig1}}
\end{figure}

The embedding geometry of the D7 probe brane in the above background (\ref{metric}) is explained in Fig.\ref{fig1}.
In general, the D7 brane bends in the transverse direction $x^{8,9}$, and this is all the data one has to specify for the shape
assuming that the shape is invariant under rotations in $x^{4,5,6,7}$ space.
Introducing the radius of $x^{4,5,6,7}$ space that the D7 brane spans
\be
\rho=\sqrt{(x^4)^2+\cdots+(x^7)^2}\quad,
\ee
the D7 brane wraps a 3-sphere of constant $\rho$ for each $\rho$. 
Note that $\rho=\rho(r,x^\mu)$ is in general a non-trivial function of holographic 5-dimensions $(r,x^\mu)$ depending on the bending in $x^{8,9}$ directions by the relation
\be
\rho=\sqrt{r^2-(X^8(r,x^\mu))^2-(X^9(r,x^\mu))^2}\quad,
\ee
where $X^{8,9}(r,x^\mu)$ are functions that specify the bending shape in $x^{8,9}$ space.
One can conveniently 
choose the eight-dimensional D7 brane world-volume coordinates to be $(x^\mu,r,\Omega_3)$, and the 
embedding geometry is completely determined by two functions $X^{8,9}(r,x^\mu)$.

The rotational $U(1)$ in $x^{8,9}$ space is a part of $SO(6)$-symmetry of the total transverse space $x^{4-9}$ 
which is the holographic manifestation of $SO(6)_R$ R-symmetry of N=4 SYM theory. For the hypermultiplets
it also corresponds to a phase rotation of the N=2 supersymmetric complex mass term, as the $r\to\infty$ asymptotic value
of $(X^8+i X^9)$ is precisely such mass parameter for the hypermultiplet. Although the hypermultiplet involves scalars too,
at least for the (Dirac) fermions it is very similar to the axial $U(1)_A$ in real QCD. We therefore call it either $U(1)_R$ or $U(1)_A$
interchangeably.  
The novelty here is that $U(1)_A$ is geometrically realized as a real rotation in $x^{8,9}$ internal space.
Several previous studies have explored
this aspect to get useful results relevant for chiral symmetry breaking phase transitions in QCD \cite{Babington:2003vm,Kruczenski:2003be,Bak:2004nt,Albash:2006ew} \footnote{As in QCD, $U(1)_A$ is not a true symmetry due to the anomaly involving gluons. A typical justification for discussing chiral symmetry breaking with it is a large $N_c$ suppression of this anomaly.}. More relevant to our present work, Ref.\cite{Hoyos:2011us} recently simulated the axial chemical potential in QCD by introducing an external time-dependent rotation
along this $x^{8,9}$ $U(1)$ angle.
One drawback is that this R-symmetry is shared by adjoint scalars and fermions in N=4 SYM theory, so that the
total charge can be lost  to the background geometry \cite{Hoyos:2011us}. However, the effective chemical potential that they introduce by rotation can be kept stationary
by continuous external inflow of necessary charges and one can meaningfully discuss the physics of hypermultiplet sector with a finite R-symmetry chemical potential. 

This bears some similarity to real QCD where axial charges may be lost due to QCD sphalerons, and indeed a time-dependent external $\theta_{QCD}$-angle (which is equivalent to the axial phase by anomaly) was one of the ways to introduce an effective axial chemical potential \cite{Kharzeev:2009fn}. It is perhaps relevant to point out that one can do a similar thing
even in the Sakai-Sugimoto model. The $U(1)_A$ suffers non-conservation due to coupling to the $C_{1}^{RR}$ which is dual to $\theta_{QCD}$-angle.  This is a holographic manifestation of $U(1)_A$ anomaly with QCD gluons. 
Although this effect has been neglected based on large $N_c$-suppression, one can go on to introduce a time-dependent $C_1^{RR}$
to introduce an effective axial chemical potential; this procedure would bypass the issues regarding the holographic chiral magnetic effect discussed in the Introduction.

The D7 brane dynamics also includes a baryonic $U(1)$ symmetry by a $U(1)$ gauge field residing on its world-volume. 
The situation we are going to study is the one having a constant magnetic field along, say $x^3\equiv z$ direction,
\be
F_{12}=B\quad,\quad 
\ee
and a finite chemical potential $\mu$, both of which are with respect to this baryonic symmetry.
Note that $B$ which is constant everywhere is a trivial solution of equations of motion.
We will focus on the massless case which means
\be
(X^8,X^9)\to 0\quad, \quad r\to \infty\quad,
\ee
and we will discuss what might be happening in the massive case in the last section.
The phase diagram of the system in $(T,B,\mu)$ has been studied previously in Ref.\cite{Evans:2010iy}, but as discussed in the Introduction
an interesting role played by $U(1)^2U(1)_R$ triangle anomaly represented by a Wess-Zumino term was overlooked.
Given $(B,\mu)$, the triangle anomaly dictates an existence of the chiral separation effect: 
\be
j_R^3 = {N_c\over 2\pi^2}\mu B\quad,
\ee
\begin{figure}[t]
	\centering
	\includegraphics[width=15cm,height=12cm]{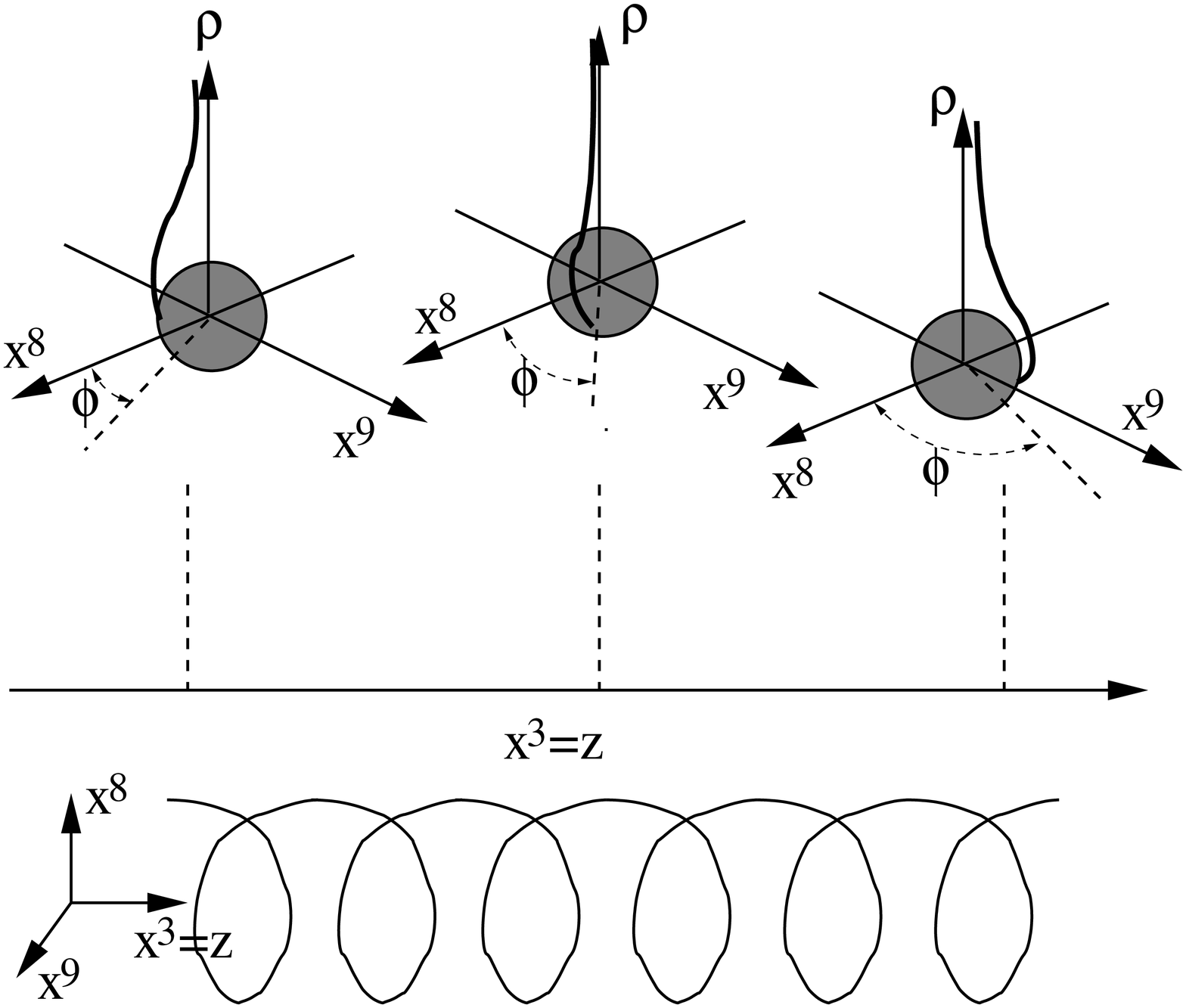}
\caption{The D7 brane embedding in the chiral helix phase. The $U(1)_R$ angle $\phi$ in $(x^8,x^9)$ plane is monotonically increasing along $x^3=z$ direction. In the space of $(x^8,x^9,z)$ the shape indeed looks like a helix.\label{fig3}}
\end{figure}
and in our case of $U(1)_R$ geometrically realized as a rotation in $x^{8,9}$ space, 
a finite $j^3_R$ current would take a form of a helix, i.e. a non-zero spatial $z$-gradient of the 
$U(1)_R$ angle of the D7 brane embedding, see Fig.\ref{fig3} for an illustration. This is an analog of the 
chiral spiral of pion gradient in low-energy QCD with electromagnetic $B$ and $\mu_B$ \cite{Son:2007ny}, except that
the axial phase and the spiral in our case are realized geometrically and are easily visualized.

This implies that a sizeable fraction of the phase diagram with $(B,\mu)$ should in fact be 
the chiral helix phase. To establish its location in the $(T,B,\mu)$ phase diagram, one should compare the grand canonical free energies (including the effects from the WZ term) of the  
phases with and without the chiral helix.
This interesting task will be pursued elsewhere \cite{col}, but in this paper we will prove the existence of this phase
by showing a dynamical instability towards it from the phase which does not possess the chiral helix.
\begin{figure}[t]
	\centering
	\includegraphics[width=10cm]{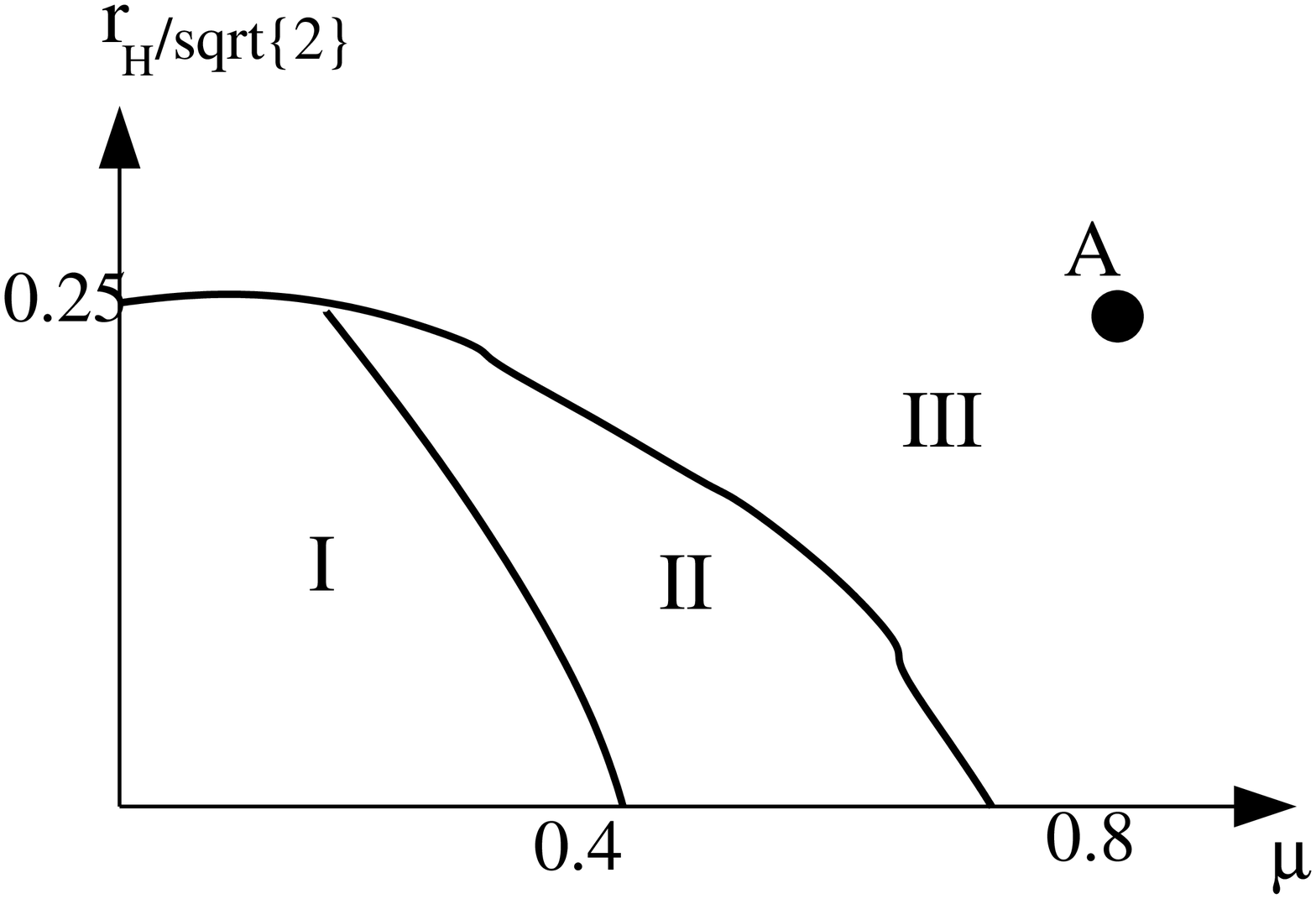}
		\caption{A schematic picture of $(T,B,\mu)$ phase diagram without considering triangle anomaly (WZ term) from Ref.\cite{Evans:2010iy,Evans:2011mu}. $B$ has been set to 1. The point A has $({r_H\over\sqrt{2}},\mu)=(0.25,1)$
and has an instability toward chiral helix phase. \label{fig4}}
\end{figure}
Fig.\ref{fig4} is a rough picture of the phase diagram of $(T,B,\mu)$ in Ref.\cite{Evans:2010iy} without considering the WZ term.
Due to conformal symmetry, the only meaningful parameters are $({\mu\over\sqrt{B}},{T\over\sqrt{B}})$ and one sets $B=1$ \footnote{
Precisely speaking there is also a rescaling $(2\pi l_s^2)F \to F$ and we choose $l_s$ to have $L^4=\lambda l_s^4 \equiv 1$ before
setting $B=1$. We also will use this convention later.}. The vertical
axis is related to $T$ by
\be
{r_H\over\sqrt{2}}={\pi\over \sqrt{2}} T\quad.
\ee
The region III in the upper-right part is the so-called supersymmetric embedding phase where
the D7 brane embedding is straight in $x^{8,9}$ space without bending all the way to the black hole horizon, $(X^8(r),X^9(r))\equiv(0,0)$. We are going to study the dynamical instability of this phase towards forming the chiral helix, and  
as an exemplar case we will pick one point -- the point A in Fig.\ref{fig4} -- which has $({r_H\over \sqrt{2}},\mu)=(0.25,1)$
and is well inside the region III without any ambiguity.
We will indeed find that this point is unstable to linearized chiral helix modes.

The action of D7 brane is
\bear
S_{D7}=-\mu_7\int d^8\xi\, e^{-\phi}\sqrt{{\rm det}(g_*+2\pi l_s^2 F)}+\mu_7 {(2\pi l_s^2)^2\over 2!}\int\, C_4^{RR}\wedge F\wedge F\quad,\label{d7action}
\eear
where $\mu_7=(2\pi)^{-7} l_s^{-8}$, $e^{\phi}=g_s$, and $F$ is the field strength of the baryonic $U(1)$ gauge field.
Note that the embedding dynamics of $X^{8,9}(r,x^\mu)$ enters through the induced world-volume metric $g_*$.
Because $l_s$ drops in any field theory observables, one can conveniently choose it such that $L^4=\lambda l_s^4\equiv 1$,
and we also rescale $(2\pi l_s^2)F \to F$ for simplicity.
We are going to study linearized fluctuations from the configuration of point A in Fig.\ref{fig4} which has $X^{8,9}(r,x^\mu)\equiv 0$. The chemical potential, or equivalently the background solution of $F_{tr}$ is
obtained from the action
\be
S_{D7}=-{\lambda N_c\over 16\pi^4}\int d^5x\, \left\{ r \sqrt{r^4+B^2} \sqrt{1-\left(F_{tr}\right)^2}\right\}\quad,
\ee
which gives one in $A_r=0$ gauge,
\be
F_{tr}^{(0)} =-\partial_r A_t^{(0)}= -{Q\over\sqrt{Q^2+r^2(r^4+B^2)}}\quad,
\ee
where a constant of motion $Q$ is determined from $\mu$ by the condition
\be
\mu=A_t^{(0)}(\infty) = \int_{r_H}^\infty dr\, {Q\over\sqrt{Q^2+r^2(r^4+B^2)}}\quad.
\ee
It is tedious but straightforward to expand the action (\ref{d7action}) quadratically from the background solution in terms of small linearized
perturbations of the modes $(X^{8,9},A_t,A_z)$ (we omit $\delta$-symbol for simplicity) assuming the dependence only on $(t,z,r)$ that are potentially
relevant for the chiral helix instability. We have verified the consistency of this ansatz. We find that
gauge field perturbations decouple from those of $X^{8,9}$ so from now on we keep only $X^{8,9}$ perturbations that are of interest for us.
One subtlety regarding $C_4^{RR}$ should be mentioned: 
writing the 5-sphere metric as
\be
d\Omega_5^2=d\chi^2+\sin^2\chi d\phi^2 +\cos^2\chi d\Omega_3^2\quad,
\ee
where $0\le\chi\le {\pi\over 2}$ is the angle from the $\rho$ axis such that (see Fig.\ref{fig1})
\be
\cos\chi={\rho\over r}\quad,\quad \sin\chi = {R\over r}\quad,\quad R\equiv \sqrt{(x^8)^2+(x^9)^2}\quad,
\ee
and $\phi$ is the $U(1)_R$ angle in $x^{8,9}$-plane, the volume form takes a form
\be
\epsilon_5=\sin\chi\cos^3\chi \,\,d\chi\wedge d\phi\wedge\epsilon_3=d\left[-{1\over 4}\cos^4\chi \wedge d\phi\wedge\epsilon_3\right]\quad,
\ee
where $\epsilon_3$ is the volume form of unit 3-sphere. From this one obtains $C_4^{RR}$ as
\bear
C_4^{RR} = {(2\pi l_s)^4 N_c\over\pi^3}{1\over 4}\left(C-\cos^4\chi\right)\wedge d\phi\wedge \epsilon_3\quad,
\eear
where a constant $C$ is a freedom of sigular gauge transformations.
Since the background D7 brane shape $X^{8,9}(r)\equiv 0$ corresponds to $\chi=0$ and we are looking at small
fluctuations around it, we need to choose $C$ such that $C_4^{RR}$ is regular around $\chi=0$. 
At $\chi=0$, the angle $\phi$ becomes singular, and $C_4^{RR}$ should vanish to be regular, which fixes $C=1$.
(The other choice $C=0$ would make $C_4^{RR}$ regular at $\chi={\pi\over 2}$ instead where the 3-sphere vanishes.
This will be suitable when discussing fluctuations around Minkowski embeddings.)
This finally gives us an expression for the WZ term:
\be
S_{WZ}={\lambda N_c\over 128\pi^4} \int d^5x\,\left(2r^2-(X^8)^2-(X^9)^2\over r^4\right)\epsilon^{MNPQR} 
\left(X^8\partial_M X^9-X^9\partial_M X^8\right) F_{NP}F_{QR}\quad,
\ee
where $\epsilon^{MNPQR}$ is purely numerical.

After a sizable amount of algebra, the quadratic action for the fluctuations one gets is 
\bear
S^{(2)}&=& {\lambda N_c\over 16\pi^4}\int d^5 x\,\Bigg\{
{1\over 2}A(r)\left(\partial_t X^8\right)^2-{1\over 2} B(r)\left(\partial_z X^8\right)^2
-{1\over 2}C(r)\left(\left(\partial_r X^8\right)^2-\partial_r\left((X^8)^2\over r\right)\right)\nonumber\\
&+& {1\over 2}D(r)(X^8)^2 + ({\rm same\,\,with\,\,} X^8\leftrightarrow X^9 ) -{1\over 2}E(r) \left(
X^8\partial_z X^9-X^9\partial_z X^8\right) \Bigg\}\quad,\label{quad2}
\eear
where the coefficient functions are given by 
\bear
A(r)&=& r\sqrt{r^4+B^2} {1\over r^4 V(r)} {1\over \sqrt{1-\left(F_{tr}^{(0)}\right)^2}} \quad,\quad B(r)=r\sqrt{r^4+B^2} {1\over r^4} \sqrt{1-\left(F_{tr}^{(0)}\right)^2}\quad,\nonumber\\
C(r)&=& r\sqrt{r^4+B^2} {V(r)\over \sqrt{1-\left(F_{tr}^{(0)}\right)^2}}\quad,\quad
D(r)=r\sqrt{r^4+B^2} {3\over r^2} \sqrt{1-\left(F_{tr}^{(0)}\right)^2}\quad,\nonumber\\
E(r)&=& {4B\over r^2} F_{tr}^{(0)}\quad.
\eear
The last term results from the WZ term which plays an essential role in our discussion.

It is a straightforward exercise to derive the equations of motion from the above and to study the linearized stability of the system.
As one has a black hole horizon located at $r=r_H$, one needs to impose a physically relevant in-coming boundary condition at
the horizon, and for this purpose it is more convenient to work in the Eddington-Finkelstein coordinate $(t_*,r_*)$,
\bear
 t_*= t+\int^r_\infty \, {dr'\over (r')^2 V(r')}\quad,\quad r_*=r\quad,
\eear
upon which one has
\bear
\partial_t=\partial_{t_*}\quad,\quad \partial_r = \partial_{r_*} +{1\over r^2 V(r)} \partial_{t_*}\quad,\quad
\int dtdr=\int dt_* dr_*\quad.
\eear
The quadratic action (\ref{quad2}) then takes a form in Eddington-Finkelstein coordinate (we omit subscript $*$ for simplicity),
\bear
S^{(2)}_{EF}&=& {\lambda N_c\over 16\pi^4}\int d^5 x\,\Bigg\{
-{1\over 2} B(r)\left(\partial_z X^8\right)^2
-{1\over 2}C(r)\left(\partial_r X^8\right)^2 -{C(r)\over r^2 V(r)}(\partial_t X^8)(\partial_r X^8)\nonumber\\
&-&{1\over 2}\left({1\over r}(\partial_r C(r))-D(r)\right)(X^8)^2 + ({\rm same\,\,with\,\,} X^8\leftrightarrow X^9 )\nonumber\\& -&{1\over 2}E(r) \left(
X^8\partial_z X^9-X^9\partial_z X^8\right) \Bigg\}\quad,\label{quad2EF}
\eear
Note that $(\partial_t X ^{8,9}) ^2$ terms disappear completely due to the identity 
\bear
A(r) = {C(r)\over r^4 (V(r))^2}\quad,
\eear  
and all the coefficient functions that appear in the above, especially $C(r)\over r^2 V(r)$, are regular at the horizon $r=r_H$.

Recall that the usefulness of the Eddington-Finkelstein coordinate is that simple regularity at the horizon $r=r_H$ automatically
guarantees the in-coming boundary condition, while out-going modes look singular in the coordinate.
An easier way to see this is to consider a wave oscillating in time $t_*$ in the Eddington-Finkelstein coordinate
\bear
e^{-i\omega t_*} = e^{-i\omega t} e^{-i\omega\int^r_\infty \, {dr'\over (r')^2 V(r')}} \quad, 
\eear
which automatically contains the necessary in-coming radial phase part in terms of the original Schwarzschild coordinate,
so that any regular wave in Eddington-Finkelstein coordinate is in-coming at the horizon.  

Equations of motion from (\ref{quad2EF}) are
\bear
&&B(r)\partial_z^2 X^{8,9}+\partial_r(C(r)\partial_r X^{8,9})+{C(r)\over r^2 V(r)}\partial_t\partial_r X^{8,9}
+\partial_t\partial_r\left({C(r)\over r^2 V(r)}X^{8,9}\right)\nonumber\\
 &&-\left({1\over r}\left(\partial_r C(r)\right)-D(r)\right) X^{8,9}
\mp E(r)\partial_z X^{9,8} =0\quad,
\eear
where the last term mixes $X^8$ and $X^9$ in such a way that the correct diagonal basis is in fact a helical basis $X^{(\pm)}=X^8\pm i X^9$, which is also tied to the $z$-direction flipping as one expects. One can easily see that
the modes at hand indeed correspond to the chiral helix modes with finite $z$-momentum.
One needs to consider only $X^{(+)}\equiv X$ in the analysis as the equations are invariant under $X^{(\pm)}\to X^{(\mp)}$ and 
$z\to-z$.

To study 
instability associated with a finite $z$-momentum one goes to frequency-momentum space by assuming $e^{-i\omega t +i k z}$,
after which one has
\bear
&&C(r)\partial_r^2 X +\left((\partial_r C(r)) -{2 i \omega}{ C(r)\over r^2 V(r)}\right) \partial_r X\nonumber\\
&&-\left({1\over r}(\partial_r C(r)) -D(r)+k^2 B(r) + k E(r) +i\omega \partial_r\left(C(r)\over r^2 V(r)\right)\right) X =0\quad.
\label{master}
\eear
As mentioned before, one imposes regularity at the horizon $r=r_H$ for in-coming boundary condition. Note that $C(r_H)=0$ and the boundary condition is not trivial. On the UV boundary $r\to\infty$, one can have two possible modes from the above equation as usual in AdS/CFT,
\bear
X\sim X_0-i\omega  {X_0\over r}+{X_1\over r^2}+\cdots\quad,
\eear
where $X_0$ corresponds to the bare mass of the hypermultiplet and the normalizability condition puts the constraint $X_0=0$
in our massless case. These two boundary conditions give a discrete spectrum of $\omega$ for a given $k$, which is an example
of quasi-normal mode problem. If the lowest $\omega(k)\to 0$ in the limit $k\to 0$, this is sometimes called hydrodynamic dispersion relation for an obvious reason, but here we are interested in the finite $k$ spectrum.
An inspection of our master equation (\ref{master}) shows that $\omega$ is purely imaginary, so that one can write
\be
\omega=i {\rm Im}[\omega]\quad.
\ee
Considering $e^{-i\omega t}$ dependence, any positive ${\rm Im}[\omega]>0$ signals an exponentially growing amplitude
 and hence an instability, while one expects ${\rm Im}[\omega]<0$ for typical dissipative relaxations.

Before delving into numerical searches for instability that will be described shortly, one can qualitatively understand from (\ref{master}) how the 
unstable modes of ${\rm Im}[\omega]>0$ can possibly appear. Consider first a fictitious situation where 
there was no WZ term, or equivalently let $E(r)$ be absent in (\ref{master}). This would be a typical situation of
D7 brane world-volume fluctuations along transverse $X^{8,9}$ directions, and one can expect that the modes would simply be
dissipating in the presence of black hole. The dependence on $k$ should obviously be that the larger $k^2$ is the faster the mode decays, so that the quasi-normal spectrum would be
\be
{\rm Im}[\omega] = -m^2_{eff}-\alpha k^2+\cdots\quad,\quad \alpha>0\quad,
\ee
for a reasonable range of $k$.
Then, in (\ref{master}) the effect of having WZ contribution is simply replacing $B(r_{eff})k^2$ with $B(r_{eff})k^2+E(r_{eff})k$ where $r_{eff}$ is the most relevant value of $r$ for the mode wave function. 
Therefore the quasi-normal mode spectrum including WZ contribution should qualitatively look like
\be
{\rm Im}[\omega] = -m^2_{eff}-\alpha k^2-\beta k +\cdots \quad,\quad \alpha>0\quad, 
\ee
where $\beta$ is roughly proportional to $E(r_{eff})$.
Completing square of the expression, one arrives at
\be
{\rm Im}[\omega] = -\alpha \left(k+{\beta\over 2\alpha}\right)^2+ {\beta^2\over 4\alpha}-m^2_{eff}=-\alpha (k-k_0)^2
+{\beta^2\over 4\alpha}-m^2_{eff} \quad,
\ee
so that when ${\beta^2\over 4\alpha}-m^2_{eff}>0$, ${\rm Im}[\omega]$ can be positive for a range of non-zero $k$ centered around $k=k_0$. This qualitative picture is confirmed  in our numerical studies. 

\begin{figure}[t]
	\centering
	\includegraphics[width=10cm]{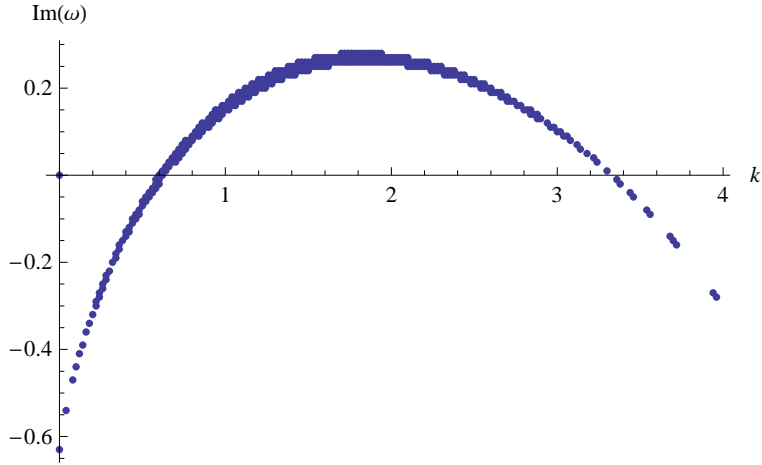}
		\caption{Our numerical result of ${\rm Im}[\omega]$ as a function of $z$-momentum $k$. We set $B=1$ and
$({r_H\over\sqrt{2}},\mu)=(0.25,1)$. This shows that  ${\rm Im}[\omega]>0$ for a range of $k$ and indicates an instability. \label{fig5}}
\end{figure}
In our numerical simulation, we solve (\ref{master}) from $r=r_H+\varepsilon$ to a UV cutoff $r=r_{max}$
given $(k,{\rm Im}[\omega])$ using a Mathematica package NDSolve.
Note that regularity at $r=r_H$ from (\ref{master}) fixes the ratios ${(\partial_r X)\over X}$ and ${(\partial_r^2 X)\over X}$
at $r=r_H$
unambiguously, which allows one to start from $r=r_H+\varepsilon$ with a small number $\varepsilon$. As the equation is linear
we are free to choose the normalization $X(r_H)=1$.
We then impose a condition of normalizability $|X(r_{max})|<\varepsilon'$ with another small number $\varepsilon'$.
We choose the parameters
\be
\varepsilon=0.001\quad,\quad r_{max}=10\quad,\quad \varepsilon'=0.05\quad,
\ee
in our numerical result in Fig.\ref{fig5}, which clearly shows a range of $k$ with positive ${\rm Im}[\omega]$.
This is a numerical proof of the existence of the chiral helix phase.

\section{Massless D3/D7 model with $(T,B,E)$}

In this section, we study another case of interest that should also be affected by the triangle anomaly, or equivalently the WZ term: the case when both electric and magnetic fields are present, and parallel to each other. The triangle anomaly dictates that
the axial $U(1)_R$ current is not conserved under this situation
\be
\partial_\mu j^\mu_R \sim \vec E\cdot \vec B\quad,
\ee
and one expects a continuous creation (or annihilation) of $U(1)_R$ charges in the system.
Noting that the $U(1)_R$ charge at hand should correspond to the angular momentum of the D7 brane along the $x^{8,9}$-plane,
the only way to satisfy this anomaly constraint is to have a spontaneous rotation of D7 brane in $x^{8,9}$ plane.
Initially, the D7 brane angular momentum would increase; then, due to the existence of black hole horizon, the 
dissipation of the created $U(1)_R$ angular momentum to the background geometry (or adjoint sector) turns on, and one can imagine a stationary situation of a constant angular momentum carried by the rotating D7 brane. We will call this a ``spontaneous rotation phase". Previous studies on the system \cite{Ammon:2009jt,Evans:2011mu} seem to have missed this possibility, and a more complete study is certainly desirable \cite{col}. In this section, we will prove the existence of this phase by showing a dynamical instability toward it from
the configuration that does not possess rotation.

\begin{figure}[t]
	\centering
	\includegraphics[width=10cm]{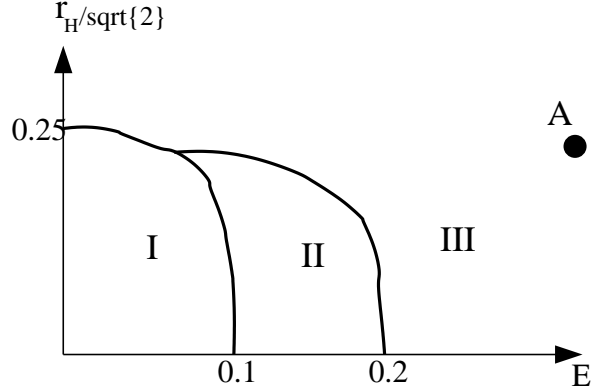}
		\caption{A rough picture of phase diagram of $(T,B,E)$ without considering triangle anomaly in Ref.\cite{Evans:2011mu}. $B$ is set to 1. The point A has $({r_H\over\sqrt{2}},E)=(0.25,1)$, which will
be shown to be unstable toward spontaneous rotation phase. \label{fig6}}
\end{figure}
The phase diagram of $(T,B,E)$ from Ref.\cite{Evans:2011mu} is sketched in Fig.\ref{fig6}. We focus only on the zero chemical potential case for
simplicity, but the essential feature of the analysis in this section is independent of the presence of chemical potential.
Using conformal symmetry, $B$ has been set to unity, and the conventions are as explained in the previous section. The region III in the far right
is again the supersymmetric embedding phase where the D7 brane is straight with $X^{8,9}(r)\equiv 0$. We will study the 
spontaneous axial rotation instability of the point A with $({r_H\over\sqrt{2}},E)=(0.25,1)$ which is well inside this region. 

To find the zeroth order background solution with constant $(B,E)$ in supersymmetric embedding phase, one starts with the relevant action component
\be
S_{D7}=-{\lambda N_c\over 16\pi^4}\int d^5x \, \left\{r\sqrt{r^4+B^2}\sqrt{1+V(r)(F_{zr})^2-{1\over r^4 V(r)}E^2}\right\}\quad,
\ee
where one turns on external magnetic $F_{12}=B$ and electric $F_{tz}=E$ fields along $z=x^3$ direction.
One can check that the constant $B$ and $E$ solve full equations of motion consistently, 
and one only needs to solve for $F_{zr}$, or $A_z$ in $A_r=0$ gauge.
Solving it from the action, one obtains 
\be
F_{zr}^{(0)} = {J\over V(r) }\sqrt{1-{1\over r^4 V(r)}E^2 \over r^2(r^4+B^2)  -{J^2\over V(r)}}\quad,\label{fzr}
\ee
where $J$ is an integration constant which is directly proportional to the induced baryonic current along $z$ direction.
One notices that the expression inside the square root in (\ref{fzr}) can change signs if $J$ is not suitably chosen \cite{Karch:2007pd}.
The on-shell action itself also involves similar factors
\be
S_{on-shell}=-{\lambda N_c\over 16\pi^4}\int d^5x\,\left\{r^2(r^4+B^2)\sqrt{1-{1\over r^4 V(r)}E^2 \over r^2(r^4+B^2)  -{J^2\over V(r)}}\right\}\quad,
\ee
and one should make sure that the total expression inside square root remains positive for a meaningful solution.
The numerator changes sign at $r=r_*$ where
\be
r_*^4=r_H^4+E^2\quad,
\ee
and this should also be the point where the denominator changes sign, which determines $J$ as \cite{Karch:2007pd}
\be
J^2=E^2\left(r_*^2+{B^2\over r_*^2}\right)=E^2\left(\sqrt{r_H^4+E^2}+{B^2\over \sqrt{r_H^4+E^2}}\right)\quad.
\ee
The point $r=r_*$ will play an important role when we discuss linearized fluctuations from this background solution.

Our task is to expand the D7 brane action quadratically for linearized fluctuations from this background solution.
After some algebra, one finally arrives at
\bear
S^{(2)} &=& {\lambda N_c\over 16\pi^4}\int d^5x\,\Bigg\{{1\over 2}A(r)\left(\partial_t X^8\right)^2
+B(r)\left(\partial_t X^8\right)\left(\partial_r X^8\right)-{1\over 2}C(r)\left(\partial_r X^8\right)^2 \nonumber\\
&-&{1\over 2}\left({1\over r}\left(\partial_r C(r)\right)-D(r)\right)(X^8)^2 +({\rm same\,\,with\,\,} X^8\leftrightarrow X^9)\nonumber\\
&+&{1\over 2}E(r)\left(X^8\partial_r X^9-X^9\partial_r X^8\right)
+{1\over 2}F(r)\left(X^8\partial_t X^9-X^9\partial_t X^8\right)\Bigg\}\quad,
\eear
where the last line is from the WZ term.
The coefficient functions are
\bear
A(r)&=& r\sqrt{r^4+B^2} {1\over r^4}\left({1\over V(r)}+\left(F_{zr}^{(0)}\right)^2\right)
{1\over \sqrt{1+V(r)\left(F_{zr}^{(0)}\right)^2 -{1\over r^4 V(r)} E^2}}\quad,\nonumber\\
B(r)&=& r\sqrt{r^4+B^2} {E\over r^4}F_{zr}^{(0)}
{1\over \sqrt{1+V(r)\left(F_{zr}^{(0)}\right)^2 -{1\over r^4 V(r)} E^2}}\quad,\nonumber\\
C(r)&=& r\sqrt{r^4+B^2} {1\over r^4}\left(r^4 V(r)-E^2\right)
{1\over \sqrt{1+V(r)\left(F_{zr}^{(0)}\right)^2 -{1\over r^4 V(r)} E^2}}\quad,\nonumber\\
D(r)&=& r\sqrt{r^4+B^2} {3\over r^2}
 \sqrt{1+V(r)\left(F_{zr}^{(0)}\right)^2 -{1\over r^4 V(r)} E^2}\quad,\nonumber\\
E(r)&=& {4B\over r^2} E\quad,\quad F(r)={4B\over r^2} F_{zr}^{(0)}\quad.
\eear
One should note that the expression 
\be
\sqrt{1+V(r)\left(F_{zr}^{(0)}\right)^2 -{1\over r^4 V(r)} E^2}=r\sqrt{r^4+B^2}\sqrt{1-{1\over r^4 V(r)}E^2 \over r^2(r^4+B^2)  -{J^2\over V(r)}}
\ee
is regular and non-vanishing at $r=r_*$ precisely because of the choice of $J$ above, so that all coefficient functions
are regular at $r=r_*$. 

It is an important fact however that $C(r_*)=0$, which implies that the regularity boundary condition at $r=r_*$ is a non-trivial one. Combined with the UV normalizability boundary condition, these two boundary conditions give us a discrete quasi-normal
spectrum of the fluctuations. Therefore $r=r_*$ (recall $r_*>r_H$), instead of the black hole horizon, seems to play a role of the IR boundary for linearized fluctuations on the D7 brane \cite{Kim:2011qh}. Because $A(r)$ remains finite at $r=r_*$, one does not 
need to work in the Eddington-Finkelstein-like coordinate around $r=r_*$, and a regular wave at $r=r_*$ is not necessarily
in-coming. 
\begin{figure}[t]
	\centering
	\includegraphics[width=10cm]{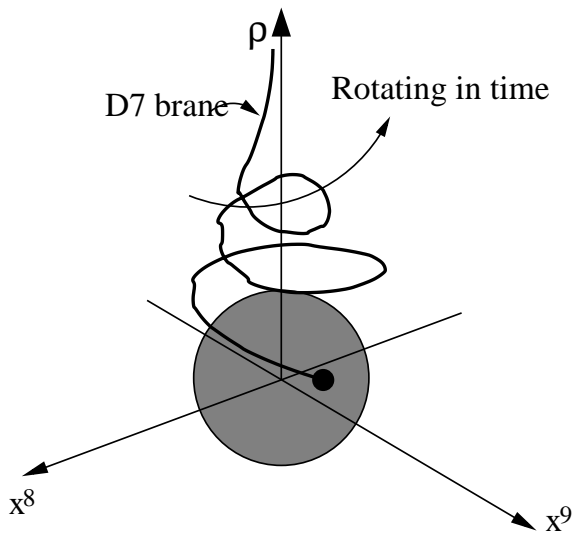}
		\caption{The shape of the D7 brane in the spontaneous rotation phase. The shape is helical along the
radial direction while the system rotates in time.\label{fig7}}
\end{figure}

The equations of motion are
\bear
&&-A(r)\partial_t^2 X^{8,9}  -B(r)\partial_t\partial_r X^{8,9}
-\partial_r\left(B(r)\partial_t X^{8,9}\right)+\partial_r\left(C(r)\partial_r X^{8,9}\right)\\
&& -\left({1\over r}\left(\partial_r C(r)\right)-D(r)\right) X^{8,9} \pm {1\over 2} E(r)\partial_r X^{9,8}
\pm{1\over 2}\partial_r\left(E(r) X^{9,8}\right) \pm F(r)\partial_t X^{9,8} =0\quad,\nonumber
\eear 
and in terms of helicity basis $X^{(\pm)}\equiv X^8\pm i X^9$, the equations of motion become diagonal.
This indicates that the D7 brane shape will be helical in the radial $r$ direction, while rotating in time, see Fig.\ref{fig7}.
To study the instability, we go to the frequency space assuming $e^{-i\omega t}$, and find that
under $X^{(+)}\leftrightarrow X^{(-)}$ the frequency maps to $\omega\leftrightarrow -\overline\omega$ so that the 
imaginary part of $\omega$ is the same for both $X^{(\pm)}$. Since the instability is signaled by a positive imaginary part ${\rm Im}[\omega]>0$, one needs to study $X^{(+)}\equiv X$ mode only. The equation to solve in frequency space is then
\bear
&&C(r)\partial_r^2 X +\Bigg(\left(\partial_r C(r)\right) +2i\omega B(r)-iE(r)\Bigg)\partial_r X \label{num2}\\
&&+\left(\omega^2 A(r)+i\omega\left(\partial_r B(r)\right)-{1\over r}\left(\partial_r C(r)\right)+D(r)-{i\over 2}\left(\partial_r E(r)\right)-\omega F(r)\right) X=0\,.\nonumber
\eear
\begin{figure}[t]
	\centering
	\includegraphics[width=10cm]{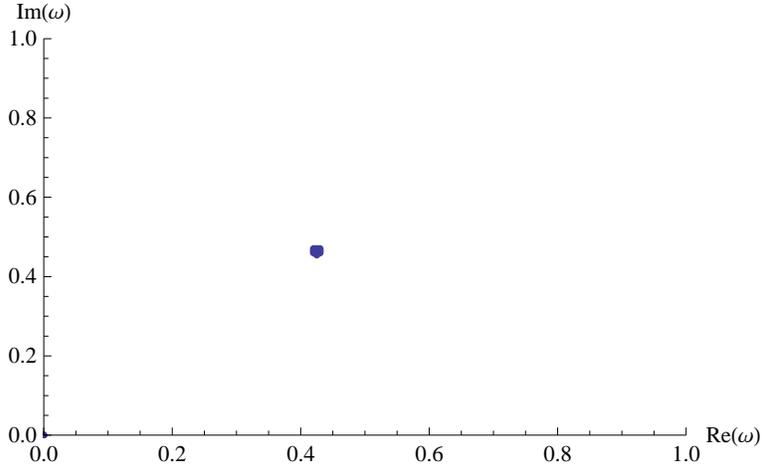}
		\caption{Numerical result for the lowest complex frequency of a linearized spontaneous rotation mode. 
The background has $({r_H\over\sqrt{2}},E,B)=(0.25,1,1)$ in the supersymmetric embedding phase. This is a numerical proof
of the existence of spontaneous rotation phase. \label{fig8}}
\end{figure}

We solve (\ref{num2}) numerically from $r=r_*+\varepsilon$ until a UV cutoff $r_{max}$ given a complex $\omega$.
One can normalize $X(r_*)=1$ and the regularity at $r=r_*$ fixes the initial conditions.
We then test a UV normalizability condition $|X(r_{max})|<\varepsilon'$, and if the solution satisfies it we let
the program put a dot in the complex $\omega$-plane. Fig.\ref{fig8} is our result using parameters
\be
\varepsilon=0.001\quad,\quad r_{max}=10\quad,\quad \varepsilon'=0.05\quad,
\ee
which clearly proves that the quasi normal spectrum $\omega$ has a positive imaginary part, and hence instability.

\section{Summary}

To summarize, we have studied the phase diagram of the D3/D7 holographic model in two cases: 1) at finite (vector) chemical potential and in the presence of an external magnetic field, and 2) in the presence of external parallel electric and magnetic fields. We have found that the triangle anomaly represented by the Wess-Zumino-like term 
in the D7 brane probe action implies the existence of previously overlooked new phases. 

In the case 1), we have found a ``chiral helix" phase 
in which the $U(1)_A$ angle of D7 brane embedding
increases monotonically along the direction of the magnetic field. We consider this as a holographic realization of the chiral spiral phase in QCD. We have found an axial current propagating in this phase, corresponding to the chiral separation effect. Previously, the existence of the chiral magnetic current at finite axial chemical potential has been established within the same D3/D7 model \cite{Hoyos:2011us}. It was argued in Ref. \cite{Kharzeev:2010gd,Burnier:2011bf} that the coupling between the axial and vector charge oscillations induced by the anomaly should lead to the emergence of a gapless excitation -- the chiral magnetic wave. It would be interesting to establish the presence of this excitation and to study its properties  within the D3/D7 holographic model.

In the case 2), we have identified the ``spontaneous rotation" phase in which the 
D7 brane spontaneously begins to rotate, so that the $U(1)_A$ angle changes as a function of time. This phase is a geometrical realization of a phase with non-zero chiral chemical potential -- indeed, the parallel external electric and magnetic fields generate chirality through the triangle anomaly.  
In the D3/D7 model the $U(1)_A$ symmetry is shared by other adjoint matter fields, so the $U(1)_A$ charges in the hypermultiplet
can be lost to the adjoint sector. This loss of chirality eventually leads to a stationary rotation frequency of the D7 brane, corresponding to some limiting value of the chiral chemical potential that can be maintained in the system. 
In real QCD plasma, the axial charge can be lost due to the sphaleron transitions; it would be interesting to establish whether or not there exists a corresponding limiting value of the chiral chemical potential similar to the one we observe in the D3/D7 model. In general, our study indicates an important role played by the anomaly in the phase diagram of gauge theories.

\vskip 1cm \centerline{\large \bf Acknowledgement} \vskip 0.5cm

The work of  D.K.\ was supported in part by the U.S.\ Department of Energy under Contracts No. DE-AC02-98CH10886 and DE-FG-88ER41723.
The work of H.U.Y. was supported by the U.S. Department of Energy under Contract No.~DE-FG02-88ER40388.

 \vfil

\end{document}